\documentclass[twocolumn,showpacs,preprintnumbers,amsmath,amssymb,aps]{revtex4}
\usepackage{graphicx}% Include figure files
\usepackage{dcolumn}% Align table columns on decimal point
\usepackage{bm}% bold math

\begin{document}

%\bibliographystyle{unsrt}

%\title{Subnatural linewidth in a strongly driven two-level system}
\title{Subnatural linewidth in a strongly-driven closed $F \rightarrow F'$ transition}
 \author{Sapam Ranjita Chanu, Alok K. Singh, Boris Brun, Kanhaiya
 Pandey, and Vasant Natarajan}
 \email{vasant@physics.iisc.ernet.in}
 \homepage{www.physics.iisc.ernet.in/~vasant}
 \affiliation{Department of Physics, Indian Institute of
 Science, Bangalore 560\,012, INDIA}

\begin{abstract}
We observe linewidths below the natural linewidth for a
probe laser on a {\it two-level system}, when the same
transition is driven by a strong control laser. We take
advantage of the fact that each level is made of multiple
magnetic sublevels, and use the phenomenon of
electromagnetically induced transparency or absorption in
multilevel systems. Optical pumping by the control laser
redistributes the population so that only a few sublevels
contribute to the probe absorption. We observe more than a
factor of 3 reduction in linewidth in the $D_2$ line of Rb
in room-temperature vapor. The observations can be
understood from a density-matrix analysis of the sublevel
structure.
\end{abstract}


%{42.50.Gy}{Effects of atomic coherence on propagation,
%absorption, and amplification of light; electromagnetically
%induced transparency and absorption}
%{32.80.Qk}{Coherent control of atomic interactions with photons}
%{32.80.Xx}{Level crossing and optical pumping}

\maketitle

The natural linewidth of a two-level transition appears as
a fundamental limit to the accuracy with which the
transition can be used, either for frequency measurement or
as a frequency reference for locking a laser. The natural
linewidth is determined by the (inverse of the) lifetime of
the excited level. In three-level and other multilevel
systems, it is now well known that a strong control laser
on an auxiliary transition can be used to modify the
lifetime of the excited level; hence the linewidth of the
transition being probed by a weak laser can become {\it
subnatural}. The effect, called electromagnetically induced
transparency (EIT) or absorption (EIA) \cite{TEA86,BIH91},
arises due to the AC-Stark shift of the levels caused by
the control laser (creation of {\it dressed states}
\cite{COR77}), and subsequent quantum interference of the
absorption pathways to these dressed states \cite{LIX95}.
The subnatural linewidth observed in these multilevel
systems \cite{RWN03,IKN08} leads to applications in
high-resolution spectroscopy \cite{RAN02} and sub-Doppler
laser cooling \cite{MOA07}, while the anomalous dispersion
near the resonance has applications in slowing of light
\cite{HHD99} and quantum information processing.

In this work, we show that such subnatural linewidth can
also be observed in a {\it two-level system}, i.e., a
system where both the control laser and the probe laser
drive the same transition. Of course, this is a two-level
system only in the sense that there is no additional level
involved and that the lifetime of the upper level limits
the linewidth of the transition. We take advantage of the
fact that each level is made of multiple magnetic
sublevels. Optical pumping by the control laser then causes
the population to redistribute among the sublevels, and
only a few sublevels contribute to probe absorption. The
coherences induced by the control laser cause a subnatural
resonance at line center. Our experiments are done in the
$5S_{1/2} \rightarrow 5P_{3/2}$ $D_2$ line of Rb, where we
observe more than a factor of three reduction below the
natural linewidth. Interestingly, we observe these narrow
resonances with room temperature vapor, where the Doppler
width is typically $100 \times$ larger than the natural
linewidth.

The ground state of Rb (and other alkali atoms) consists of
two hyperfine levels, thus there are two sets of
transitions starting from this state. Though the subnatural
feature shows up for both sets of transitions, it appears
as enhanced absorption (EIA) in one case and enhanced
transmission (EIT) in the other. This is because the
dominant transition for the lower-level set is the closed
$F \rightarrow F-1$ transition (with a smaller number of
magnetic sublevels in the excited state), while it is $F
\rightarrow F+1$ for the upper-level set (with a larger
number of magnetic sublevels in the excited state). The
differences between these two sets and the observed line
shapes can be understood from a density-matrix analysis of
the system.

The subnatural features we observe in our work are also to
be contrasted with the phenomenon of coherent-population
trapping (CPT) in three-level $\Lambda$ systems
\cite{AGM76,ARI96}, where the linewidth can be extremely
narrow compared to the linewidth of the excited state
because it is limited only by the decoherence rate between
the two ground levels. This phenomenon relies on the use of
{\it phase-coherent} control and probe lasers to drive the
atoms into a dark non-absorbing state. The two lasers have
to have roughly equal intensities as both are required to
drive the atoms into the dark state. Since CPT is a
ground-state coherence phenomenon, it is used, for example,
for precision spectroscopy of the ground hyperfine interval
in atomic clocks \cite{WYN99}. These experiments benefit by
the use of vapor cells filled with buffer gas or with
paraffin coating on the walls, which increase the
ground-coherence time but also shift or broaden the optical
transition. Moreover, the relevant natural linewidth in
these cases is the inverse of the lifetime of the upper
ground level, which can be quite long because the
transition is electric-dipole forbidden. The scan axis in
CPT experiments is the Raman detuning between the two
lasers from the two-photon resonance condition, and not the
optical frequency of the probe laser. Thus, if the control
laser were detuned from its optical resonance, it will not
affect the Raman resonance condition for CPT, but it will
shift the resonance location for EIT. In summary, our EIT
experiments (i) use a weak probe laser; (ii) can be used
for precision spectroscopy on the excited state
\cite{RAN02}; (iii) use pure vapor cells; and (iv) have a
scan axis (showing the subnatural feature) that is the
optical frequency of the {\it phase-independent} probe
laser.

\begin{figure}
\centering{\resizebox{0.95\columnwidth}{!}{\includegraphics{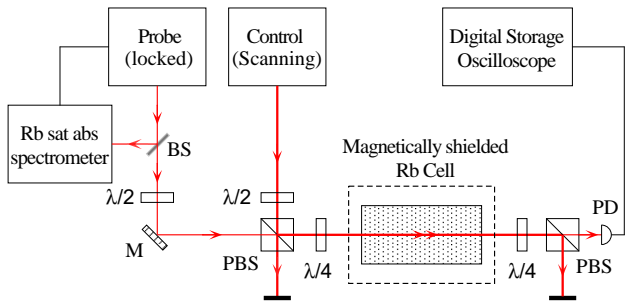}}}
\caption{(Color online) Schematic of the experiment with
independent control and probe lasers. For the experiments
with phase coherent beams, both beams were derived from the
same laser. Figure key: BS -- beamplitter, $\lambda/2$ --
halfwave retardation plate, $\lambda/4$ -- quarterwave
retardation plate, M -- mirror, PBS -- polarizing
beamsplitter cube, PD -- photodiode.}
 \label{schema}
\end{figure}

The experimental set-up is shown schematically in Fig.\
\ref{schema}. The control and probe beams are derived from
two {\it independent} home-built diode laser systems
operating on the 780 nm $D_2$ line of $^{87}$Rb
\cite{BRW01}. The linewidth of the lasers after feedback
stabilization is about 1 MHz. The beams are elliptic and
have a size of 2 mm $\times$ 3 mm. The probe beam is locked
to a hyperfine transition using saturated-absorption
spectroscopy (SAS) in a vapor cell. The control beam is
scanned around the same transition. The two beams have
orthogonal circular polarizations ($\sigma^+$ and
$\sigma^-$) and co-propagate through a vapor cell. The cell
has a multilayer magnetic shield that reduces the stray
fields to below 1 mG. The two laser beams are mixed and
separated using polarizing beam splitter cubes (PBS), and
the probe beam is detected with a photodiode. The
individual beam powers are controlled using halfwave
retardation plates before the PBS's. We have shown earlier
\cite{BAN03} that the use of co-propagating beams
eliminates crossover resonances, and that by scanning the
control laser while keeping the probe fixed makes the
signal appear on a Doppler-free background.

\begin{figure}
\centering{\resizebox{0.75\columnwidth}{!}{\includegraphics{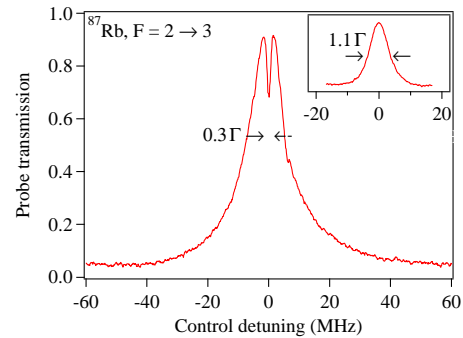}}}
\caption{(Color online) Subnatural EIA resonance for
upper-level transitions obtained with the probe laser
locked to the $F=2 \rightarrow F'=3$ transition and control
laser scanning across the same transition. The inset shows
a typical SAS spectrum, which has a linewidth of $1.1
\Gamma$.}
 \label{2to3}
\end{figure}

The first set of experiments was done for transitions
starting from the upper ground level in $^{87}$Rb, i.e., on
the closed $F=2 \rightarrow F'=3$ transition. The results
for a probe power of 8 $\mu$W and control power of 150
$\mu$W are shown in Fig.\ \ref{2to3}. The spectrum shows a
broad ($\approx 20$ MHz wide) transparency peak as the
control is scanned. This is the usual resonance occurring
due to EIT and saturation effects seen in pump-probe
spectroscopy \cite{BAN03}. Exactly at line center, a narrow
EIA dip (corresponding to enhanced absorption) appears. The
full-width-at-half-maximum (FWHM) of the narrow resonance
is only 1.8 MHz compared to the natural linewidth of 6 MHz.
The subnatural feature is robust and the FWHM remains less
than 3 MHz ($0.5 \Gamma$) with increase in control power to
250 $\mu$W. But as the control power is increased, the
depth of the resonance and its signal-to-noise ratio
increases. For comparison, a typical SAS spectrum (taken
with counter-propagating beams) is shown in the figure
inset. With optimal powers in the pump and probe beams, the
linewidth is 6.6 MHz ($1.1 \Gamma$).

\begin{figure}
(a)\centering{\resizebox{0.95\columnwidth}{!}{\includegraphics{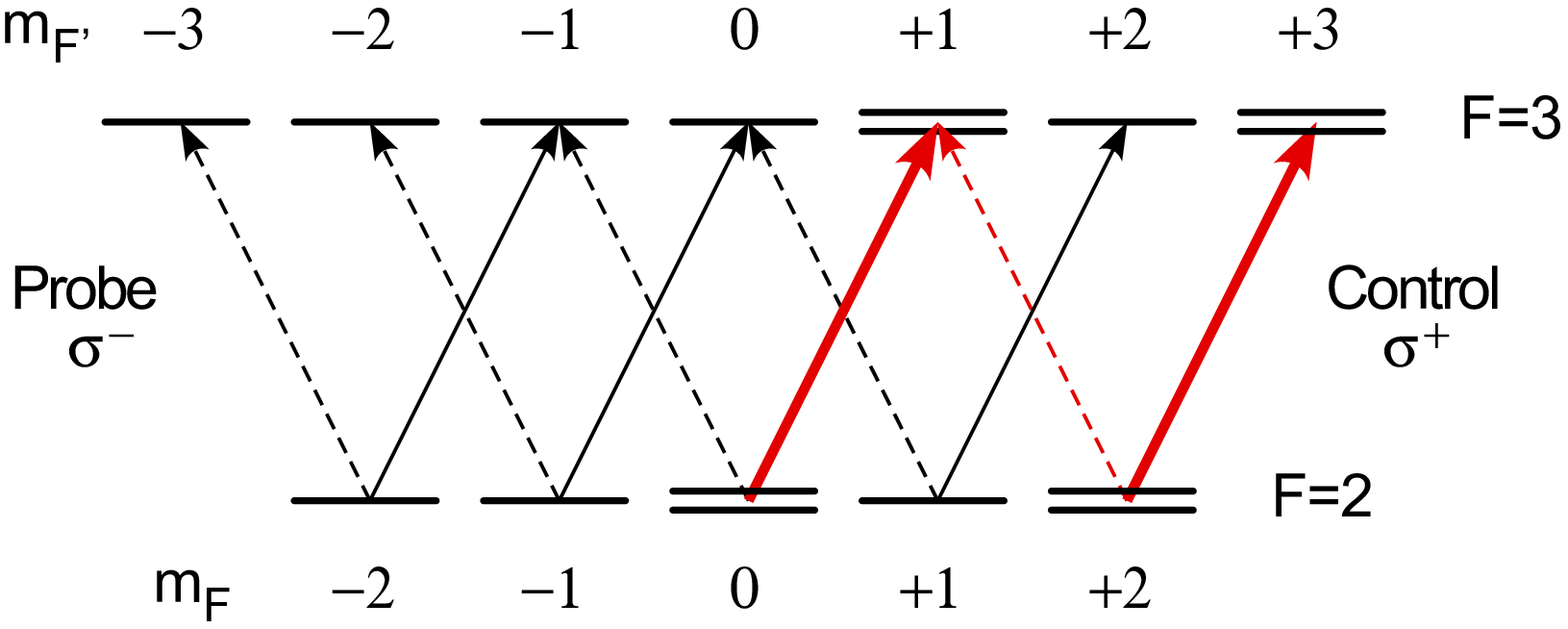}}}
(b)\centering{\resizebox{0.75\columnwidth}{!}{\includegraphics{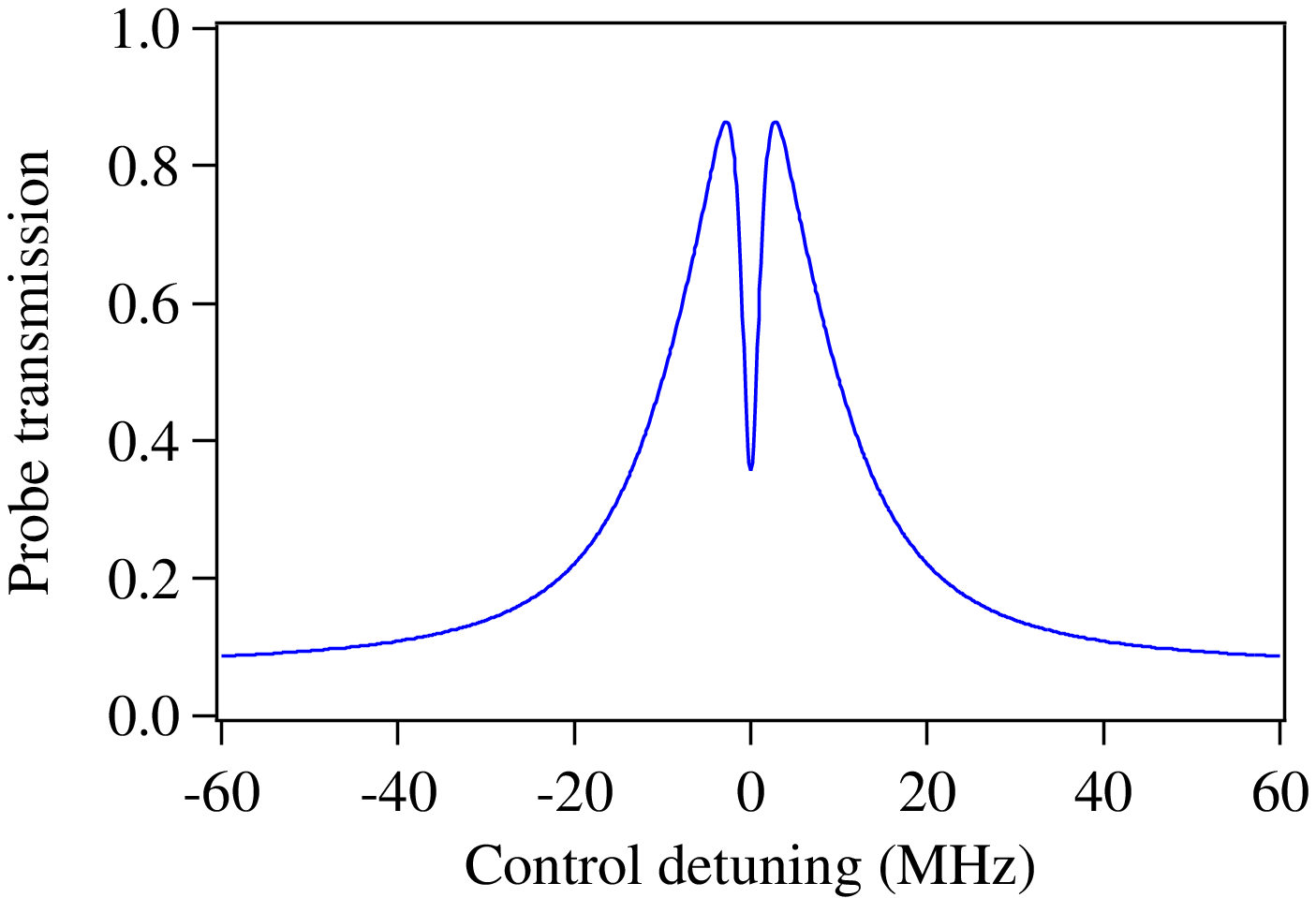}}}
\caption{(Color online) Theoretical study of upper-level
transitions. (a) Sublevel structure for the $F=2
\rightarrow F'=3$ transition. Optical pumping causes only
the $m_F=0$ and $m_F=+2$ sublevels to contribute. (b)
Calculated spectrum for the $N$-type system in (a).}
 \label{Ntype}
\end{figure}

To understand this theoretically, we consider the magnetic
sublevel structure shown in Fig.\ \ref{Ntype}. For the $2
\rightarrow 3$ transition, there are 5 and 7 sublevels
respectively. Optical pumping by the circularly-polarized
control will transfer all the population into the $m_F=+2$
ground sublevel, as shown in Fig.\ \ref{Ntype}(a). If we
therefore ignore the other sublevels, we have a V-type
system formed by the $m_{F'}=+1 \leftrightarrow m_{F}=+2
\leftrightarrow m_{F'}=+3$ sublevels. Dressing of the
$m_F=+2$ and $m_{F'}=+3$ sublevels by the strong control
laser will cause the usual transparency resonance (EIT) for
the weak probe laser. Now consider that the control is also
going to dress the $m_F=0$ and $m_{F'}=+1$ sublevels,
forming effectively an $N$-type system \cite{GWR04,BMW09}.
The additional coherences induced by this is what causes
the narrow EIA resonance.

We have done a standard density-matrix analysis of this
system to confirm the above explanation. The calculation
takes into account Doppler averaging in room-temperature
vapor. In Fig.\ \ref{Ntype}(b), we show the calculated
spectrum considering the $N$-type system formed by the
$m_F=0$ and $+2$ sublevels. The Rabi frequency of the
control laser is taken to be 6 MHz, which corresponds to
the experimental power if we assume that the entire power
is spread uniformly over the beam size. To account for the
finite linewidth of the control laser, we assume that the
ground sublevels decohere at a rate of 1 MHz. There are no
other adjustable parameters. The calculated spectrum
reproduces the features of the observed spectrum shown in
Fig.\ \ref{2to3}, with a broad EIT peak and a narrow EIA
dip at line center. We have verified that the EIA dip
disappears if we do not consider the dressing of the
$m_{F'}=+1$ sublevel. Furthermore, the depth of the EIA
resonance increases with increasing Rabi frequency of the
control, exactly as observed experimentally. The
calculation also shows that the width of the EIA dip
becomes smaller at smaller control powers, but observing
linewidths below 1.5 MHz is experimentally challenging
because the probe laser itself has a linewidth of 1 MHz.

\begin{figure}
(a)\centering{\resizebox{0.75\columnwidth}{!}{\includegraphics{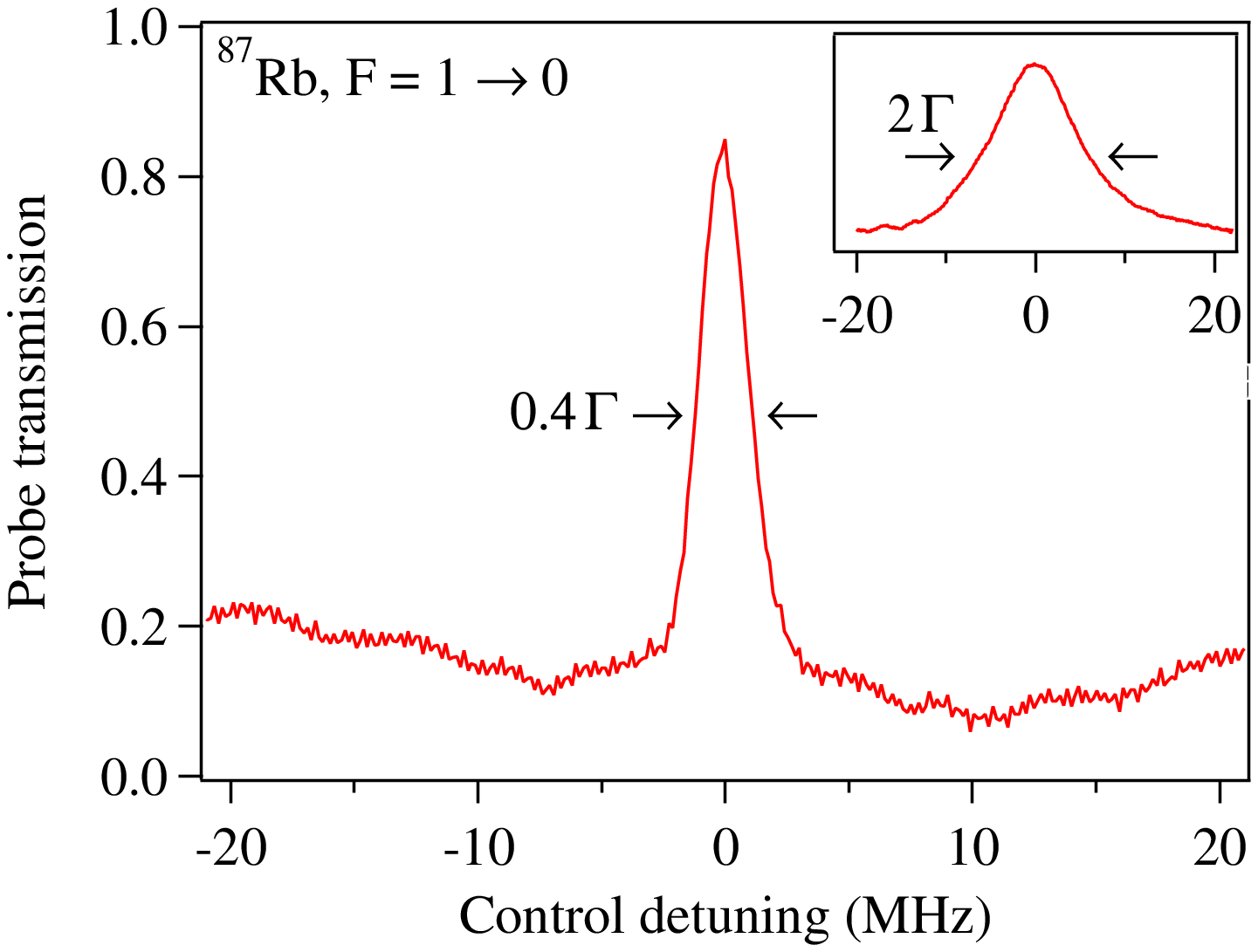}}}
(b)\centering{\resizebox{0.75\columnwidth}{!}{\includegraphics{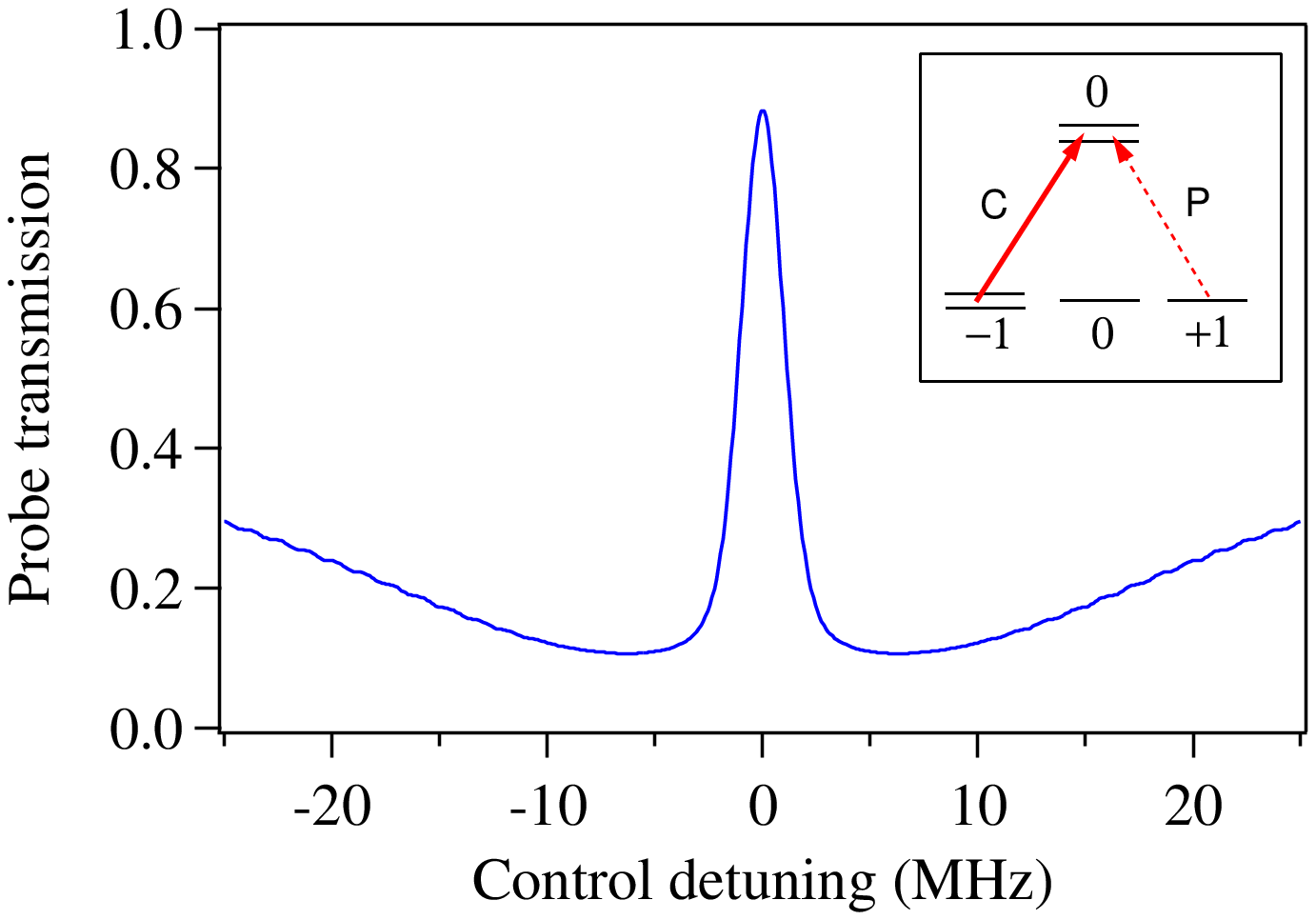}}}
\caption{(Color online) Subnatural EIT resonance for
lower-level transitions. (a) Measured spectrum with the
probe laser locked to the $F=1 \rightarrow F'=0$
transition. The inset shows a typical SAS spectrum for the
same transition. (b) Calculated spectrum for the
$\Lambda$-type system shown in the inset.}
 \label{1to0}
\end{figure}

The next set of experiments was done for transitions
starting from the lower ground level in $^{87}$Rb, i.e.,
with the lasers on the closed $F=1 \rightarrow F'=0$
transition. The spectrum obtained with a probe power of 8
$\mu$W and a smaller (compared to the previous set) control
power of 80 $\mu$W is shown in Fig.\ \ref{1to0}(a). In this
case, the measured spectrum has a central {\it
transparency} peak surrounded by broad enhanced absorption
wings. The EIT peak at line center is {\it subnatural},
with a linewidth of only 2.4 MHz ($0.4 \Gamma$. For
comparison, we again show in the inset a typical SAS
spectrum taken with optimal pump and probe powers. The
linewidth of 12 MHz is 5 times larger than the one for the
``controlled'' case.

As before, the features of the measured spectrum can be
understood theoretically. In this case, the sublevel
structure is quite simple, with 3 sublevels in the lower
level and 1 sublevel in the upper level. The
circularly-polarized beams couple these sublevels to form a
$\Lambda$-type system, as shown in the inset of Fig.\
\ref{1to0}(b). The calculated spectrum, using a Rabi
frequency of 4.4 MHz (because the control power is a factor
of 2 smaller) and taking into account thermal averaging,
reproduces the features of the observed spectrum including
the enhanced-absorption wings. This can be understood as
follows. Off resonance, the strong control laser optically
pumps the population into the $m_F=+1$ sublevel, which
causes increased probe absorption. As the control comes
into resonance, there are additional induced coherences,
which causes the EIT resonance. It is well known that in
$\Lambda$ systems the EIT resonance can be subnatural
\cite{LIX95,RWN03}. The calculated spectrum has a width
similar to the measured one.

As in the case of upper-level transitions, the FWHM of the
central resonance remains below 3 MHz even with a control
power of 250 $\mu$W, while its signal-to-noise ratio
increases. We are also able to go to lower control powers
of about 40 $\mu$W and still see a prominent resonance. But
the linewidth does not decrease much, probably again
because of the 1-MHz linewidth of the probe laser.

\begin{figure}
(a)\centering{\resizebox{0.75\columnwidth}{!}{\includegraphics{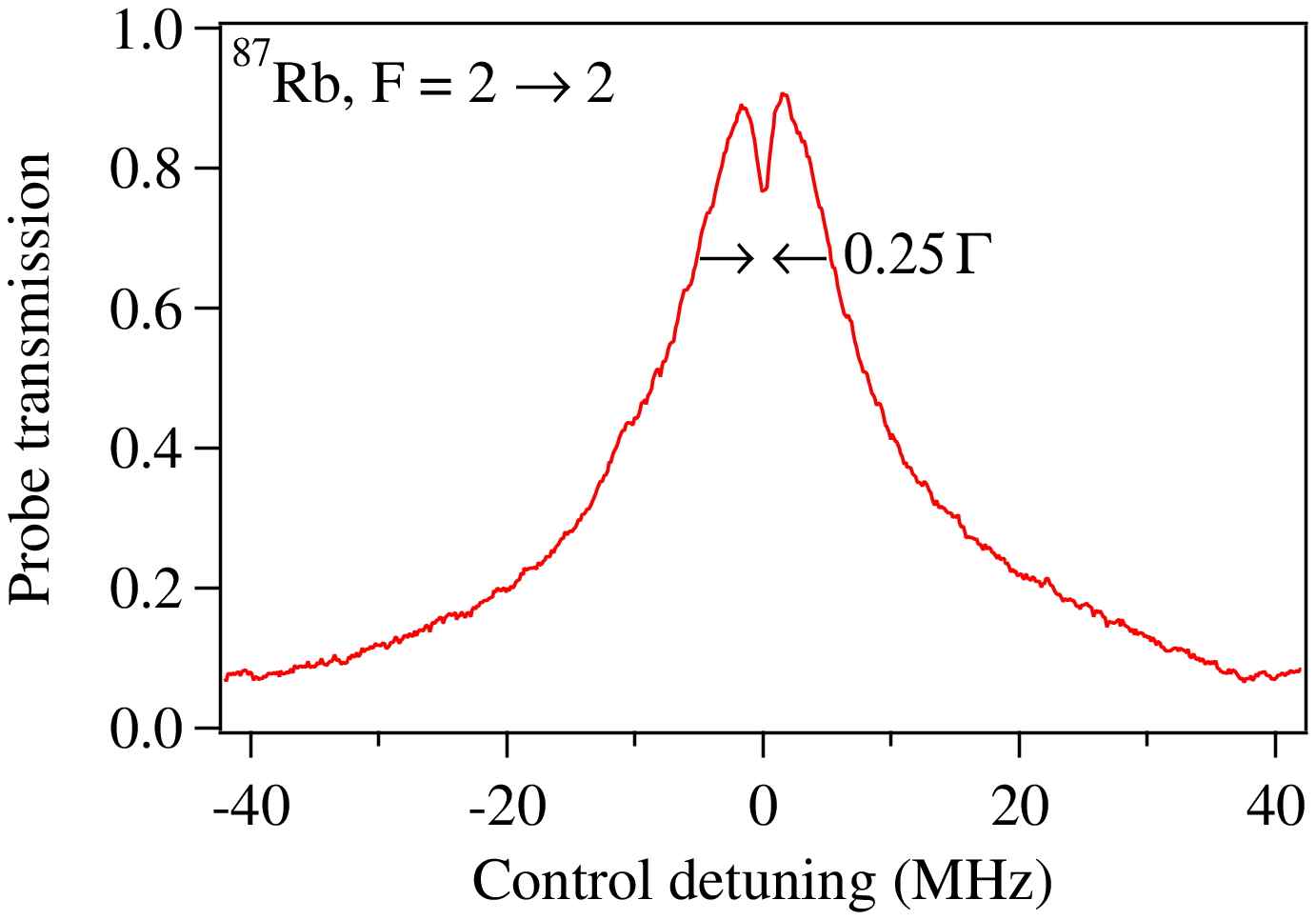}}}
(b)\centering{\resizebox{0.75\columnwidth}{!}{\includegraphics{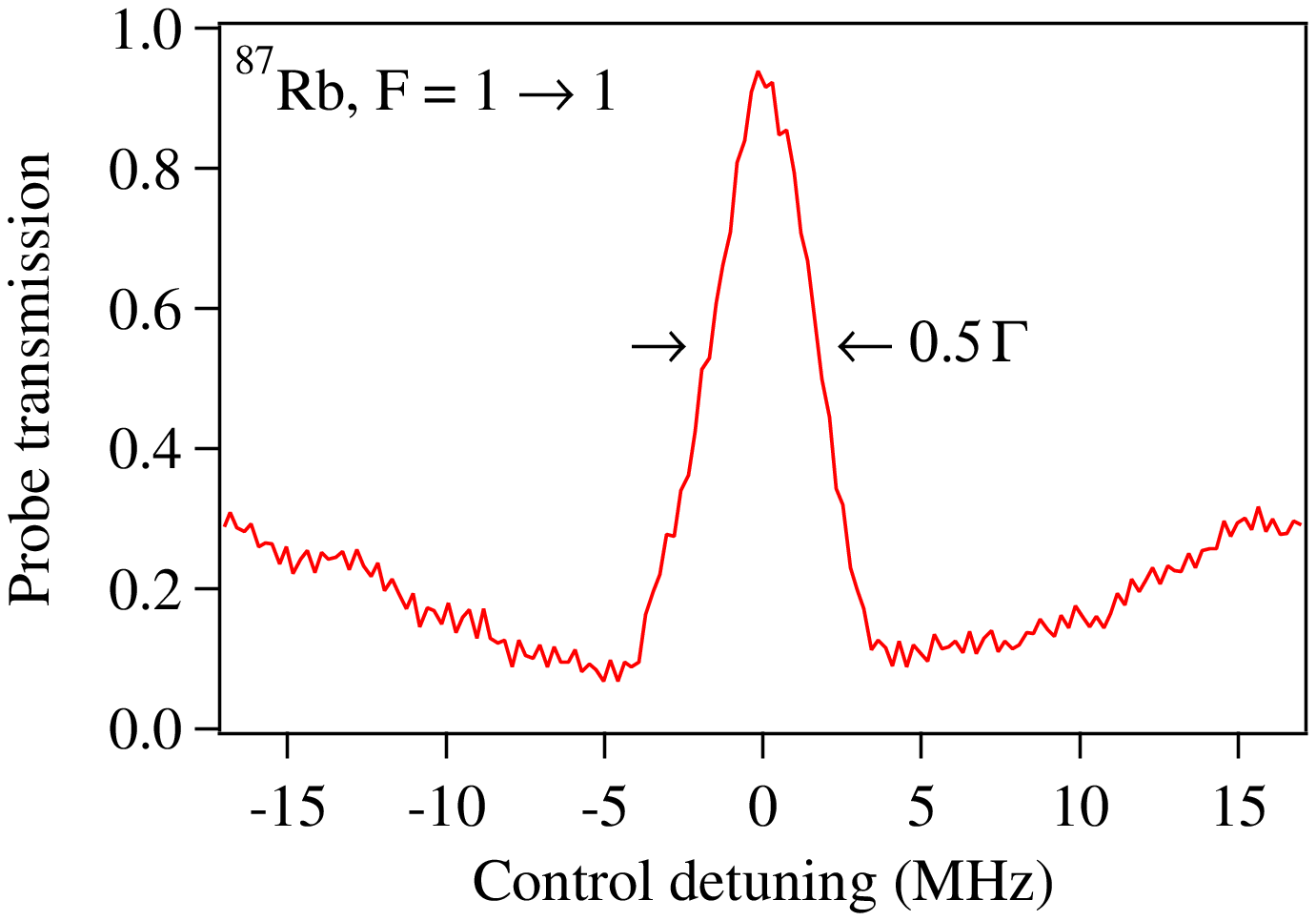}}}
\caption{(Color online) Subnatural resonances for the open
transitions. (a) is for upper-level transitions showing
enhanced absorption, while (b) is for lower-level
transitions showing enhanced transparency. The widths are
similar to what is observed for the closed transitions.}
 \label{open}
\end{figure}

In the above experiments, we have considered the closed
transition in each set, partly because it is a better
realization of a two-level system, and partly because we
can do a more rigorous theoretical calculation without
worrying about the presence of the other levels. This is
not the case for open transitions because we have to
consider the effect of optical pumping into the closed
transition. However, experimentally we find the same
behavior for the open transitions also. The results for the
$2 \rightarrow 2$ and $1 \rightarrow 1$ transitions are
shown in Fig.\ \ref{open}. The control powers are 150
$\mu$W and 40 $\mu$W, respectively. As before, we see
enhanced absorption for the upper-level transition [shown
in (a)] and enhanced transmission for the lower-level
transition [shown in (b)]. The line shapes and widths are
similar to the closed transitions, because the closed
transition is the dominant transition in each set and takes
over after a few absorption-emission cycles.

We thus see that the subnatural feature is quite robust,
appearing for both closed and open transitions, and for
control powers ranging from 5 to 30 times the probe power.
It appears even with linear polarization of the two beams
and does not seem to require circular polarization. In
addition, it appears without the magnetic shield around the
cell so that the sublevels are not degenerate. Finally,
since our cell contains both isotopes of Rb, we can do the
same experiments with the other isotope, namely $^{85}$Rb.
We observe the same behavior, i.e., an enhanced-absorption
subnatural resonance for upper-level transitions ($F=3
\rightarrow F'$) and an enhanced-transmission subnatural
resonance for lower-level transitions ($F=2 \rightarrow
F'$).

In conclusion, we have observed linewidth reduction below
the natural linewidth in a two-level system when a strong
control laser is applied to the same transition. We take
advantage of the presence of multiple magnetic sublevels in
each level and the phenomenon of electromagnetically
induced transparency or absorption, a phenomenon that is
well studied in multilevel systems. In this effect, the
control laser causes population redistribution and
``dressing'' of the sublevels, and the linewidth reduction
comes about because of interference among the absorption
pathways. We observe these subnatural resonances in the
$D_2$ line of $^{87}$Rb using room-temperature atoms in a
vapor cell. There are two sets of hyperfine transitions
starting from the two ground hyperfine levels, and we
observe the subnatural feature for both sets. However, it
appears as an enhanced-absorption dip for upper-level
transitions, and as an enhanced-transparency peak for
lower-level transitions. This difference can be understood
from the differences in the number of magnetic sublevels
for the dominant closed transition in each set. The
observed line shapes and widths are reproduced by
density-matrix calculations. We observe more than a factor
of 3 reduction in linewidth below the natural linewidth,
while conventional saturated-absorption spectroscopy gives
linewidths that are 1 to 2 times the natural linewidth.
These subnatural features and the consequent anomalous
dispersion should prove useful in high-resolution
spectroscopy and other applications of EIT since they
appear exactly at line center.

This work was supported by the Department of Science and
Technology, India. V.N. acknowledges support from the Homi
Bhabha Fellowship Council; and A.K.S. and K.P. from the
Council of Scientific and Industrial Research, India.

%\bibliography{D:/papers/eitsubnat/eitrefs}

\end{document}